\documentclass[11pt,a4paper]{iopart}
\pdfoutput=1
\usepackage{iopams}
\usepackage{graphicx,epsfig}
\usepackage{dcolumn}
\usepackage{bm}
\usepackage[applemac]{inputenc}
\usepackage{hyperref}
\expandafter\let\csname equation*\endcsname\relax
\expandafter\let\csname endequation*\endcsname\relax
\usepackage{amsmath}
\usepackage{color}

\newcommand{\Msol}{\ensuremath{M_{\odot}}}

\newcommand{\deltatomega}{\Delta \mathfrak{T}_{\rm SZ}}

\newcommand{\aap}{A \& A}
\newcommand{\apj}{ApJ}
\newcommand{\apjs}{ApJS}
\newcommand{ \mnras}{MNRAS}
\newcommand{\jcap}{JCAP}
\newcommand{\aaps}{A\&AS}

\begin{document}

\title[Constraining the CMB temperature evolution with SZ spectra]{Constraining the evolution of the CMB temperature with SZ measurements from Planck data}

\author{G. Luzzi$^{a,}$\footnote{Corresponding authors} , R. T. G\'{e}nova-Santos$^{b,}$\footnotemark[\value{footnote}], C.J.A.P. Martins$^{c,d}$, M. De Petris$^{a}$, L. Lamagna$^{a}$}
{\it 
$^a$Dept. of Physics, Sapienza, University of Rome. Piazzale Aldo Moro 2, I-00185 Rome, Italy.\\
$^b$Instituto de Astrof\'{\i}sica de Canarias, C/V\'{\i}a L\'{a}ctea s/n, La Laguna,
Tenerife, Spain.\\
$^c$Centro de Astrof\'{\i}sica, Universidade do Porto, Rua das Estrelas, 4150-762 Porto, Portugal.\\
$^d$Instituto de Astrof\'{\i}sica e Ci\^encias do Espa\c co, CAUP, Rua das Estrelas, 4150-762 Porto, Portugal\\}

\begin{abstract}
The CMB temperature-redshift relation, $T_{\rm CMB}$(z)=$T_0$(1+z), is a key
prediction of the standard cosmology, but is violated in many non standard
models. Constraining possible deviations to this law is an effective way to test the $\Lambda$CDM paradigm and to search for hints of new physics. We have determined $T_{\rm CMB}$(z), with a precision up to $3\%$, for a subsample (104 clusters) of the Planck SZ cluster catalog, at redshift in the range 0.01-- 0.94, using measurements of the spectrum of the Sunyaev Zel'dovich effect obtained from Planck temperature maps at frequencies from 70 to 353 GHz. The method adopted to provide individual determinations of $T_{\rm CMB}$(z) at cluster redshift relies on the use of SZ intensity change, $\Delta I_{\rm SZ}$($\nu$), at different frequencies, and on a Monte-Carlo Markov Chain approach. By applying this method to the sample of $104$ clusters, we limit possible deviations of the form $T_{\rm CMB}$(z)=$T_0$(1+z)$^{1-\beta}$ to be $\beta= 0.022 \pm 0.018$,  at 1$\sigma$ uncertainty, consistent with the prediction of the standard model. Combining these measurements with previously published results we get $\beta=0.016\pm0.012$.

\end{abstract}


\section{\label{intro} Introduction}
In the past two decades cosmology became a data-driven science, and we have learned more about the universe in this
period than in all the rest of mankind's history. The synthesis of this knowledge is provided by the so-called
concordance cosmological model, which provides a very good fit to a plethora of data with only a fairly small number
of free parameters and theoretical assumptions. Still, this comes with some cost: about $96\%$ of the contents of the
universe should be in two forms---dubbed dark matter and dark energy---that have never been detected in the
laboratory, but have only been 'statistically discovered' in astrophysical and cosmological data, through their
inferred gravitational properties.

As a matter of fact, the observational evidence for the recent acceleration of the universe demonstrates that the
canonical theories of cosmology and particle physics must be at least incomplete---if not incorrect---and that there
is new physics waiting to be discovered. The next generation of flagship astrophysical facilities must therefore be
optimized for the role of searching for, identifying and ultimately characterizing this new physics. A key component
of this roadmap is a significant improvement in the precision of consistency tests of the standard paradigm. This
paper provides a contribution along these lines.

If the expansion of the Universe is adiabatic and the cosmic microwave background (CMB) spectrum was originally a
black-body, its shape will be preserved by the subsequent evolution, with its temperature behaving as
\begin{equation}
T_{\rm CMB}(z)=T_0(1+z)\,.
\end{equation}
This is a robust prediction of standard cosmology, but it is violated in non-standard models, the most obvious
scenarios being those involving photon mixing and/or violation of photon number conservation. There are many
non-standard, but nevertheless theoretically well-motivated, physical processes in which photon number is not
conserved. Examples include a non perfectly transparent Universe, decaying vacuum cosmologies (with photon injection
mechanisms), models in which the fine-structure constant varies, physically motivated scenarios where photons mix
with other particles (such as axions), many modified gravity scenarios, and so on. Therefore, direct observational constraints on this relation lead to constraints on a range of scenarios beyond the  standard $\Lambda$CDM paradigm.
It's also worth noting that additional indirect constraints stem from the distance duality relation (as discussed
in \cite{Avgoustidis2012,Avgoustidis2014}) and from CMB spectral distortions \cite{Chluba2014}.

The COBE-FIRAS experiment has revealed a very precise black-body spectrum, with a temperature of $T_0=2.725\pm0.002$
K \cite{Mather}, but this measurement per se tells us nothing about the behaviour of the CMB at non-zero redshifts.
Currently there are two astrophysical techniques that can probe this behaviour. At low redshifts (typically $z<1$),
the CMB temperature can be measured via the thermal Sunyaev-Zel'dovich (SZ) effect towards galaxy clusters \cite{fabbri1978,reph1980}.
This method was first applied to ground-based CMB observations of a small number of clusters \cite{Battistelli,Luzzi2009},
which demonstrated its potential, and more recently it was further explored with the current generation of ground and
space experiments \cite{Avgoustidis2012,Atrio,Hurier2014,Saro2014}. Here we build upon these earlier works.

At higher redshifts (typically $z>1$), the CMB temperature can be determined from quasar absorption line spectra which
show atomic and/or ionic fine structure levels excited by the photon absorption of the CMB radiation \cite{Bahcall}.
The CMB is an important source of excitation for species with transitions in the sub-millimeter range, such as C, CO or CN. (Of these, CN is in principle the best thermometer, but so far it hasn't been detected in high-redshift systems.)
Although the concept is more than four decades old, measurements (as opposed to upper bounds) were only obtained much
more recently \cite{Srianand2000}, with errors at the ten percent level \cite{Noterdaeme2011}  and for one case at a few percent level at $z=0.89$ \cite{Muller2013}. Significant improvements are expected from ALMA \cite{ALMA1,ALMA2} and especially from the E-ELT high resolution
spectrograph \cite{EELT1,EELT2}, particularly through CO measurements which are signal-to-noise limited.

We note that the two techniques complement each other, not only in terms of redshift coverage but also because they
have  entirely different potential systematics. Spectroscopic observations probe the matter era, and despite the
shortage of  suitable targets (which is in fact the limiting factor for this method) the large redshift lever arm can
lead to significant constraints. On the other hand, SZ clusters probe the epoch of dark energy domination and its
onset, and large catalogs of SZ clusters are now available, making it a powerful probe.

In the present work we focus on a subsample (containing a total of 107 clusters) of the Planck SZ cluster catalog, in
the redshift range 0.01-- 0.94, to constrain possible deviations of the temperature-redshift relation. As is common
in the existing literature, we will phenomenologically parametrize possible deviations by
\begin{equation}
T_{\rm CMB}(z)=T_0(1+z)^{1-\beta}\ ,
\end{equation}
as proposed by \cite{Lima200}. We will therefore obtain constraints on the dimensionless parameter $\beta$, both from our cluster sample alone and
in combination with other direct measurements, improving currently existing bounds \cite{Noterdaeme2011,Avgoustidis2012,Muller2013,Hurier2014,Saro2014}.

The structure of the rest of the paper is as follows. In section~2 we describe the different datasets used to determine 
the final SZ fluxes and the cluster catalogue. In sections~3 and 4 we explain respectively the procedure to obtain the SZ fluxes and 
to estimate the $T_{\rm CMB}(z)$ temperature at the position of each cluster. In section~5 we present our final results and 
in section~6 we summarise the main conclusions of this work.

\section{\label{data} Data}
The core of the present analysis rests upon a subsample of the Planck SZ cluster
catalog (PSZ) \cite{cpp29} and Planck temperature maps at frequencies
from 70 to 857 GHz \cite{cpp1}, on the use of X-ray data from BAX (X-ray Cluster Database)
\cite{BAX} and from the MetaCatalog of X-ray detected Clusters of
galaxies (MCXC) \cite{Piffaretti2011}.

The Planck spectral coverage allows us to explore the positive and negative part of the SZ spectral distortion, it is optimally suited for cluster detection and, more important for this project, allows for breaking cluster parameters
degeneracy, when including priors over one or more of the parameters (e.g. the electronic temperature). Here we focus on a study dedicated to a sample of known clusters, for which X-ray and optical information are available, i.e. a subsample of the Planck SZ cluster catalog. Taking advantage of the multi-frequency capability, the high quality of data (S/N ratio $\ge$6) and the precise absolute calibration (known to be better than 1$\%$) we will show that it is possible to determine the CMB temperature at clusters redshift with a precision of up to $3\% $, comparable to the results based on the use of radio-mm molecular absorbers at $z=0.89$ from \cite{Muller2013}.

\subsection{Planck temperature maps}

To derive the SZ fluxes we use Planck DR1 maps \cite{cpp1}, which cover a wide frequency range delimited by nine 
independent channels, which are provided by the LFI (30, 44 and 70~GHz) and the HFI (100, 143, 217, 353, 545 and 857~GHz) 
instruments\footnote{These maps, and all the rest of Planck-related public products that have been used in this paper, have been 
downloaded from the Planck Legacy Archive,\\
 \url{http://www.sciops.esa.int/index.php?project=planck\&page=Planck\_Legacy\_Archive}}. 
 These maps are provided in the HEALPix pixelization \cite{gorski05}, using a pixel resolution of $N_{\rm side}=2048$ for the HFI 
 channels and $N_{\rm side}=1024$ for the LFI channels. All the results presented in this paper were obtained at full resolution, 
 $N_{\rm side}=2048$, for which we had to re-pixelize the LFI maps. Although we clean the foreground contributions from these 
 maps at each cluster position, to avoid contamination from residuals we ignore the extreme frequencies, and focus our analyses 
 in the five frequency bands between 70 and 353~GHz. The first three channels lie in the negative part of the SZ spectrum, the 217~GHz 
 channel is close to the null, whereas at 353~GHz the SZ effect is positive.
 
\subsection{Ancillary maps}

Ancillary data will be used to clean our maps from thermal dust emission and from the CMB fluctuations, following the methodology 
explained in section~\ref{cleaning}. As a template to correct for the dust emission around each cluster we use the $100~\mu$m IRIS 
map \cite{miville2005}. This map is a reprocessing of the original map from the Infrared Astronomical Satellite (IRAS), in which 
the calibration, zero level and striping problems are largely corrected.

To remove the emission from the CMB we resort to LGMCA (local-generalized morphological component analysis)
\footnote{Downloaded from {\tt http://www.cosmostat.org/planck\_wpr1.html}.}, a CMB map produced by 
\cite{bobin14} using a combination of Planck DR1 and WMAP 9-year data. We initially attempted to use three of the CMB maps released 
in Planck DR1, resulting from different component-separation techniques \cite{cpp12}: SMICA, NILC and SEVEM. However, as it will 
be shown in section~\ref{cleaning}, these maps suffer from important SZ contamination, so we use instead LGMCA which does not have 
this problem. Indeed the authors claim that under the assumption of a perfect instrument calibration, this CMB map has zero SZ 
residuals.

	\subsection{\label{subPSZ}Selection of the PSZ subsample}
		
	We use the Planck all-sky SZ (PSZ) cluster catalog. PSZ has 1227 entries, among these 861 are confirmed clusters, and 813 have known redshift.
	Among these 813 clusters, we have selected a sample of 107 clusters organized into 
	subsamples. A subsample of 75 clusters with $z<0.4$ have signal-to-noise threshold of 
	seven. In these clusters $R_{500}$, the radius within which the mean overdensity of the cluster is 500 times the critical density at the cluster redshift, is taken from MCXC and the electronic temperature, $T_e$, is 	given by BAX. The next subsample is made of 11 clusters 
	at intermediate redshifts ($0.3<z\le 0.6$), and also have signal-to-noise threshold of seven. Here $R_{500}$ is taken from MCXC, and $T_e$, under the assumption of isothermal gas, is estimated using the relation $\bar{T} - M_{500}$ given in \cite{Arnaud2005}:
	\begin{equation}
		E(z)M_{500}=A_{500}\left(\frac{k\bar{T}}{5keV}\right)^{\alpha_T}
	\end{equation}
	with $\alpha_T=1.49$, $A_{500}=4.10 \times 10^{14} h_{70}^{-1} \Msol$, and $M_{500}$ is taken from MCXC (there was no match with BAX). 
	We have also included in 	our sample 21 high-redshift clusters, with $ z > 0.6$ and up to z=0.972; only for clusters ACT-CLJ0102-4915 (El Gordo) and PSZ1 G266.56-27.31 we have $R_{500}$ and $T_e$ from respectively \cite{Menanteau2012} and  \cite{pip26}, otherwise for the other 19 clusters we use $R_{500}$ as estimated by Planck and $T_e$ estimated using scaling relations. In particular, to estimate $R_{500}$ when no X-ray informations 	are  available we marginalize over Y (the integrated Compton parameter within 5$R_{500}$) the two-dimensional probability distribution in $\theta_\mathrm{500}$ (the estimated size) and $Y$ given by the PWS detection algorithm, one of	the 	three detection algorithms used by the Planck collaboration to produce the union Planck SZ catalogue (see Planck release 2013). Although these 19 high-redshift clusters have quite larger uncertainties on $R_{500}$ with respect to other clusters in the subsample we have included them because of higher lever arm on $\beta$ due to the exploration of the distant universe, and also to show the power of a per cluster analysis with the Planck frequency coverage (with respect to the SPT results \cite{Saro2014}) and to compare our results to the ones obtained by \cite{Hurier2014}\footnote{They have applied a stacking of PSZ clusters in different redshift bins, with only one
	cluster in each of their highest redshift bins z = 0.8 and z = 1.} . 
	 		
	The selection of the present subsample has been motivated with the aim of using the same cluster sample, through SZ and X-ray measurements, to study not only the CMB evolution, but also the Hubble diagram and the distance 		duality relation, $\eta(z)=D_{L}(z)/[(1+z)^{2}D_{A}(z)]=1$, where the last equality holds for the standard model, but not for others. Identical sky coverage and redshift distribution will considerably reduce the impact of
	astrophysical systematics (in particular for the distance duality test) and represents a unique opportunity to check the
	consistency of those models for which the cosmological distance duality relation and the CMB temperature scaling law
	are simultaneously modified \cite{Avgoustidis2012}. 
	We will investigate the distance duality relation in a follow-up publication, in this case we will not include the 19 clusters of our subsample for which we don't have X-ray informations on $R_{500}$.

\section{\label{method_szflux}Estimation of the SZ fluxes at Planck frequencies}

	\subsection{\label{cleaning}Map cleaning}
	
In order to obtain reliable SZ fluxes we have to clean the Planck frequency maps from other contributions. In particular, the 
synchrotron and the free-free emissions, which affect mainly the low-frequency LFI data, the thermal dust emission, which is important 
at the HFI frequencies, the CMB, and extragalactic point sources. There are two further contaminants that affect the HFI maps: the 
zodiacal light, and CO emission, which affects the 100, 217 and 353~GHz channels, as a consequence of the (1-0), (2-1) and (3-2) 
rotational transition lines, respectively. For the HFI channels, we use publicly-available zodiacal-light corrected maps, which 
have been produced by analyzing data from different surveys which are subject to different levels of zodiacal light intensities as 
explained in \cite{cpp6}. The Planck collaboration has also released templates to correct for the CO emission in the 100, 217 and 
353 GHz channels \cite{cpp13}. Three different methodologies are used, leading to Type 1, 2 and 3 maps, respectively. Type 2 and 
Type 3 maps are known to have significant contamination from other foregrounds. In fact, via stacking of these maps on the positions 
of our 107 clusters we find significant SZ residuals, which are on average $\sim -7~\mu$K per cluster at 100 GHz and $\sim 35~\mu$K 
at 217 GHz on the Type 2 maps, and $\sim 25~\mu$K on the 100 GHz Type 3 map. On the other hand, Type 1 maps have no residuals, but 
suffer from too low signal-to-noise in regions away from the Galactic plane, not being useful for our purposes. Therefore, we prefer 
not to correct for the CO emission.

The contamination of extragalactic point sources is removed through the application of the {\it SZ union} mask, which is the one used 
by the Planck collaboration when building the SZ catalogue \cite{cpp29}, and is a combination of the point-source masks  
of the six HFI frequency channels. All the pixels affected by this mask are ignored in our analysis. On the other hand, the 
free-free and synchrotron emissions may affect the lowest LFI frequencies. The Planck collaboration has provided maps of the spectral 
index and amplitude of this emission at 30~GHz, which could be used to apply a pixel-by-pixel correction. However, these maps have 
an angular resolution of 30~arcmin, and are produced at a lower $N_{\rm side}$ of 256. Low-frequency surveys such as the 
Haslam map at 0.48~GHz \cite{haslam82} or the Reich map at 1.4~GHz \cite{reich86}, which are commonly used as templates to clean for this 
emission at higher frequencies, can neither be used here owing to their low angular resolution, $\sim 1^\circ$. As the median 
angular size of the clusters in our catalogue is $\theta_{500}\sim 6$~arcmin, the large-scale emission traced by these maps will be 
largely cancelled through the subtraction of the background in our aperture photometry technique. Therefore, we do not perform any 
cleaning of the synchrotron or free-free emission, and instead decide to ignore in our analysis the first two LFI bands (30 and 44~GHz), 
which present the highest contamination from these foregrounds.

\begin{figure}
	 \begin{center}
    		\includegraphics[width=16cm,keepaspectratio]{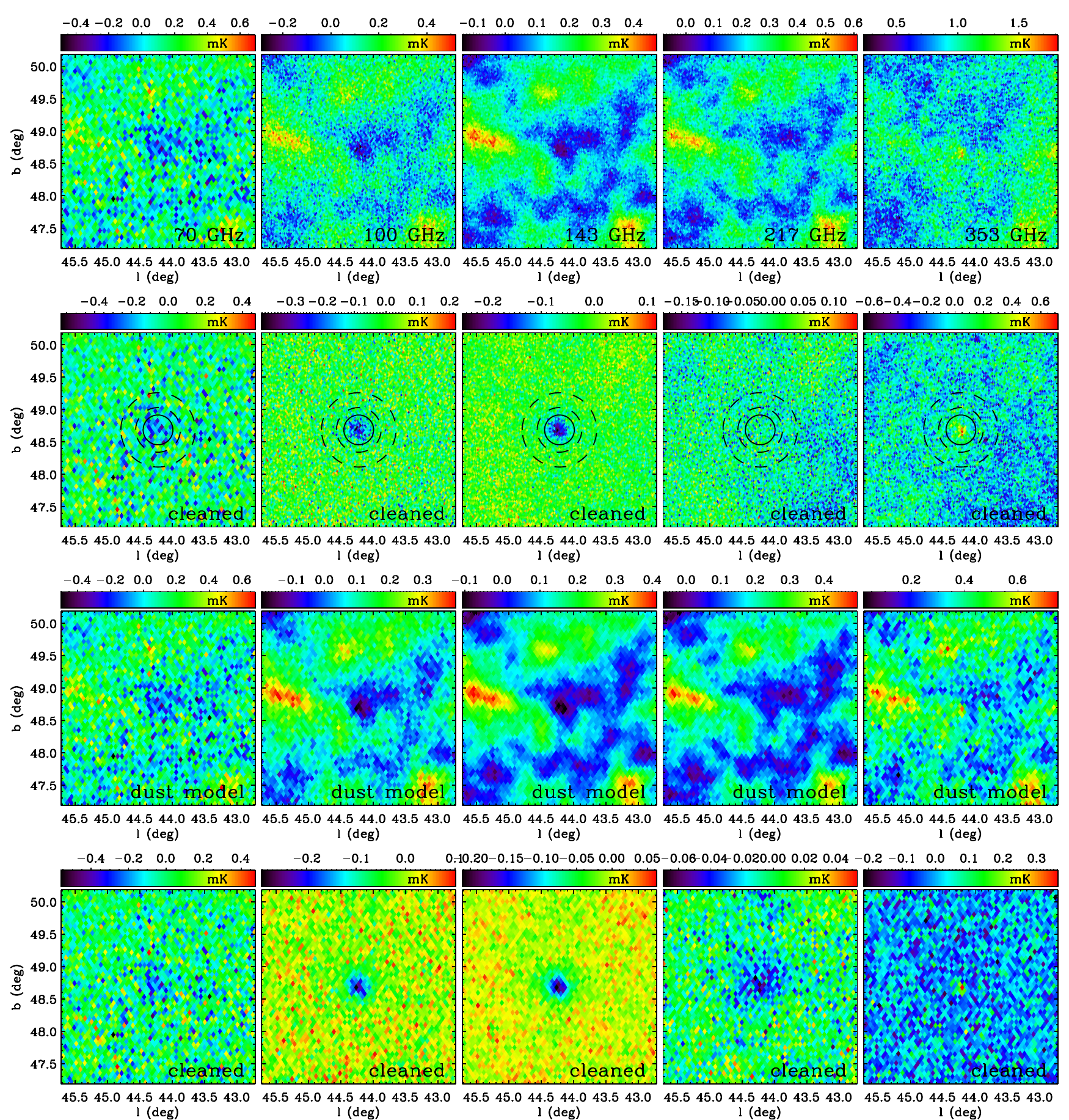}
    		\caption{\label{fig_dustcleaning} Planck LFI and HFI maps around the position of the cluster PSZ1G044.24+48.66 
    		(A2142). From left to right we show maps from 70 to 353~GHz, the five frequencies used. The first line 
    		corresponds to the original Planck frequency maps, and the second line to the maps after cleaning for the thermal 
    		dust and CMB using our methodology (see section~\ref{cleaning}). The maps clearly show the expected SZ decrements 
    		at frequencies below 217~GHz, no effect at 217~GHz, and a positive signal at 353~GHz. The solid circle depicts the 
    		aperture we use to integrate the SZ flux, and the two concentric dashed rings the annulus used for background 
    		subtraction. The third line corresponds to the maps of the Planck dust model, showing in some frequencies the SZ 
    		residuals of this model. The last line shows the maps cleaned using this model. The negative feature at 217~GHz 
    		comes from the SZ residuals, and highlights the unsuitability of these maps.}
  	 \end{center}
	\end{figure}

The thermal dust emission is the most important contaminant at HFI frequencies. One possible way for cleaning this foreground is by
using the maps of the three parameters that define the modified black-body (MBB) emission law (dust-grains temperature, emissivity 
index and optical depth) derived from the Planck dust model \cite{cpp11}, as it is done in \cite{genova14}. However, this model 
has important SZ residuals, as it is highlighted in Figure~\ref{fig_dustcleaning}, which presents the dust maps derived from 
this model between 70 and 353~GHz at the position of the cluster A2142. We can see by eye a negative SZ residual at 100 and 143 GHz. 
Although not visible, the residual at 217~GHz is positive, which becomes evident in the dust-subtracted map, where a negative signal 
appears at the position of the cluster. We have performed a stacking of these dust maps at the positions of our 107 galaxy clusters, finding 
that on average the SZ residuals are at a level of around $3$, $6$, $20$ and $120~\mu$K, respectively at frequencies 100, 143, 217 and 
353~GHz. 

As an alternative, we apply the well-known method, proposed by \cite{diego02} in the context of SZ studies, consisting on finding 
the parameter $\alpha(\nu)$ which minimises the variance of the subtraction
\begin{equation}
M_{\rm dc}(\nu,x) = M(\nu,x)-\alpha(\nu)M_{\rm d}(x)~~,
\end{equation}
where $M_{\rm d}$ is a template tracing the dust emission, and $M$ is the frequency map that we want to clean at a given 
position $x$. The value of this parameter that minimizes the variance is 
\begin{equation}
\alpha(\nu) = \frac{\sum_{i}M(\nu,x_i)M_{\rm d}(x_i)}{\sum_{i} M_{\rm d}(x_i)^2}~~,
\label{eq:alpha}
\end{equation}
where the sum extends over all the pixels of a certain sky region. Instead of assuming the same $\alpha$ value for the 
whole sky, we build cleaned mini-maps around each cluster position, and use a radius $5\theta_3$ around each cluster, 
with $\theta_3$ being the external radius of the background ring that we use in our aperture photometry technique 
(see next section). 
As dust template, $M_{\rm d}$, we first use the 100~$\mu$m IRIS \cite{miville2005} 
map\footnote{Downloaded from {\tt http://www.cesr.fr/$\sim$bernard/Ancillary/IRIS/}.}. To further reduce 
possible residuals, we perform a second iteration using the Planck 857~GHz map.

\begin{figure}
	 \begin{center}
    		\includegraphics[height=4.8in,width=5in,keepaspectratio]{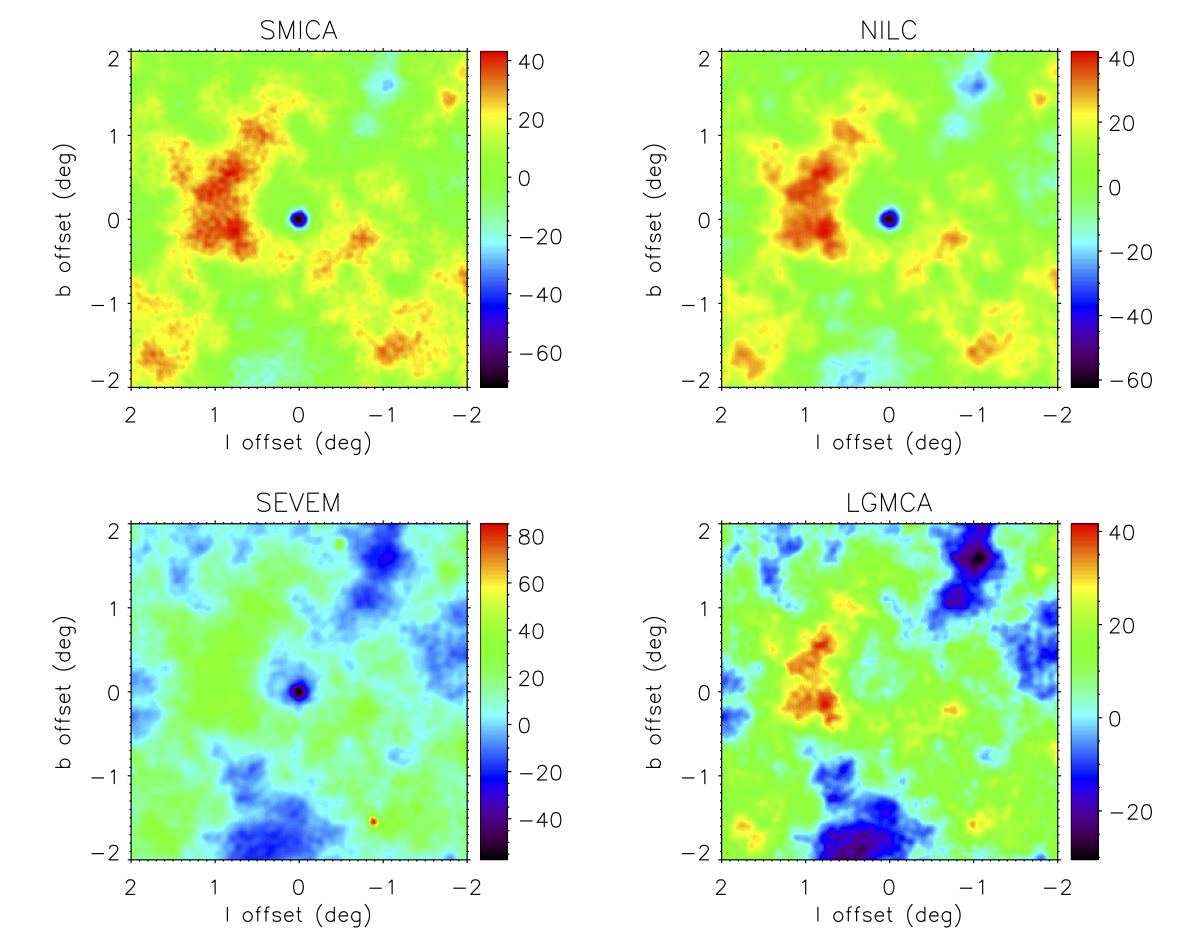}
    		\caption{\label{fig_LGMCA} 
                Stack of different CMB-reconstructed maps at the positions of the clusters of our catalogue. The maps are 4 degrees on a side. The units of the colour 
		scale are mK$_{\rm CMB}$, and the maps have been divided by 107 in such a way that each pixel represents the average of the temperatures of the 
		107 maps around each cluster in that specific position. We consider three foreground-cleaned CMB maps delivered by 
		the Planck collaboration \cite{cpp12} which are the result of independent component-separation methods (SMICA, NILC and SEVEM) and 
		also the LGMCA map produced by \cite{bobin14}. As we see, the LGMCA map is the only one showing no clear SZ residuals, and therefore we use this map 
		in our analysis.}
  	 \end{center}
	\end{figure}

The last step in our cleaning procedure consists on removing the CMB fluctuations. Whereas the 217~GHz map is commonly used as a template 
to remove the CMB in the context of SZ studies, in our case this approach would introduce important modifications in our analysis.
Including the 217~GHz map subtraction in the fitting procedure introduces a strong degeneracy between the optical depth and $T_{\rm CMB}$, thus preventing a meaningful determination of the CMB evolution.
We first tried to use 
three of the Planck DR1 CMB maps (SMICA, NILC and SEVEM), but the SZ residuals preclude their use. These are evident in 
Figure~\ref{fig_LGMCA}, where we show a stack of each of these maps in the positions of the 107 clusters of our catalogue. A 
negative feature is seen towards the centre of these maps, with a value of $\sim -40~\mu$K, that represents the average minimum 
SZ temperature decrement on these clusters. On the contrary, the same stack on the LGMCA CMB map \cite{bobin14} renders no visible 
thermal SZ residuals. Therefore, we will use the LGMCA map to subtract the CMB from each of the six frequency maps that we use. 
The side effect is that the LGMCA method takes the standard thermal SZ spectrum, without relativistic corrections and assuming the standard scaling of the CMB temperature, explicitly into account
during the component separation. This could bias our analysis. 
Assuming for example that the CMB temperature is systematically lower with respect to the standard case means that the LGMCA maps should show an excess at cluster positions. By making the stacking (Figure~\ref{fig_LGMCA}) we see that there is no such an effect, or at least it is into the noise level. Of course, this doesn't mean that we are no more sensitive to the determination of $T_{\rm CMB}(z)$, because in order to extract it we use multifrequency informations and not only the 217 GHz channel.
It is anyway possible that, in case of non negligible relativistic corrections to the thermal and kinematic SZ components, we mimic a kinematic SZ (KSZ) in our spectrum, because of the subtraction at each frequency of the same amount of signal. As will be shown in ~\autoref{mcmc} we fit for a KSZ component in our analysis. The uncertainty in the determination of $T_{\rm CMB}(z)$ takes this into account.

The final cleaned map is given by
\begin{equation}
 M_{\rm c}(\nu,x) = M(\nu,x)-\alpha_{\rm IRIS}(\nu)M_{\rm IRIS}(x)-\alpha_{\rm 857}(\nu)M_{\rm 857}(x)-M_{\rm LGMCA}(x)~~,
 \label{eq:mc}
\end{equation}
where $M$, $M_{\rm IRIS}$, $M_{857}$ and $M_{\rm LGMCA}$ are respectively the original Planck frequency map, the IRIS $100~\mu$m map, 
the Planck 857~GHz map and the LGMCA CMB map. $\alpha_{\rm IRIS}$ is obtained through equation~\ref{eq:alpha}, using as a dust 
template the IRIS map and as the data the initial frequency map. In the calculation of $\alpha_{857}$ we use the 857~GHz map as template 
and the map cleaned using IRIS as the data. 

When applying equation~\ref{eq:mc}, we take all maps to a common angular resolution. This means that for $\nu\ge 217$~GHz we degrade 
$M$, $M_{\rm IRIS}$ and $M_{857}$ to 5~arcmin, the angular resolution of the LGMCA map. The beam FWHMs for $\nu<217$~GHz are larger than 
5~arcmin; therefore, in those cases we degrade $M_{\rm IRIS}$ and $M_{857}$ and $M_{\rm LGMCA}$ to the angular resolution of $M(\nu)$.

\begin{figure}
	 \begin{center}
    		\includegraphics[width=16cm,keepaspectratio]{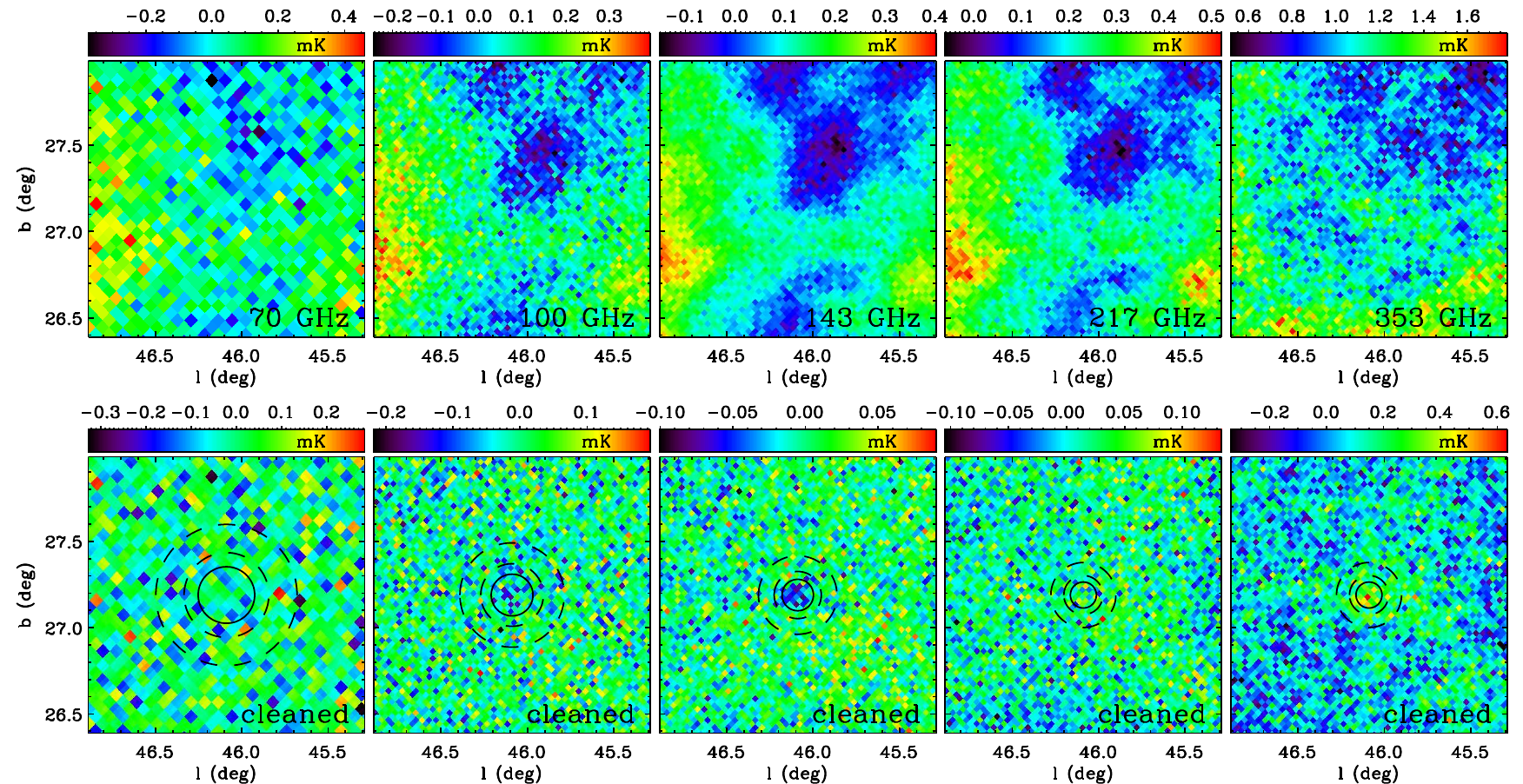}		
    		\caption{\label{fig_dustcleaning2} Planck LFI and HFI maps around the position of the cluster PSZ1G046.09+27.16. 
    		The first and the second lines show respectively the original frequency maps and the cleaned maps, as in 
    		Figure~\ref{fig_dustcleaning}. Despite the high redshift ($z=0.389$) and small angular size 
    		($\theta_{500}=4.5$~arcmin), its SZ signal is still evident at 143 and 353~GHz, proving the efficiency of our 
    		cleaning methodology.}
  	 \end{center}
	\end{figure}

In Figure~\ref{fig_dustcleaning} we show the final cleaned maps at the position 
of the galaxy cluster A2142, where the SZ effect shows up at all frequencies except 217~GHz. This is a nearby cluster at 
$z=0.0894$ and with a relatively large angular size, $\theta_{500}=13.8$~arcmin. To illustrate the ability of our cleaning 
methodology to recover the SZ signal in more distant clusters, in Figure~\ref{fig_dustcleaning2} we show the cleaned maps towards 
the cluster PSZ1G046.09+27.16, at redshift $z=0.389$ and with angular size $\theta_{500}=4.5$~arcmin. Even for this distant cluster 
the SZ signal is clearly detected at 143 and 353~GHz.
	
\subsection{\label{aph}Aperture photometry}

\begin{figure}
 \begin{center}
\includegraphics[width=16cm,keepaspectratio]{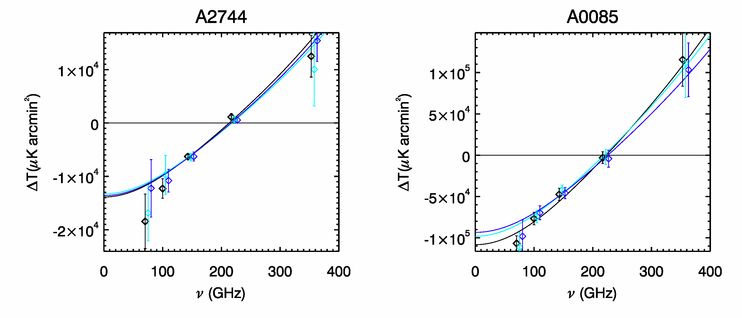}
\caption{\label{fig_3ringfluxes} 
We show the SZ fluxes derived through aperture photometry in two nearby clusters (A2744 and A0085), using three different sets of parameters 
for the radius of the aperture ($\theta_1$) and for the internal ($\theta_2$) and external ($\theta_3$) radii of the ring used for background 
subtraction. We use [$\theta_1,\theta_2,\theta_3$] = [$1.0$,$1.5$,$2.5$]$\theta_{500}$ (black points), 
[$1.0$,$2.0$,$2.24$]$\theta_{500}$ (cyan) and [$1$,$3$,$4$]$\theta_{500}$ (blue). The solid lines 
show the fitted SZ spectra for each case.
}
\end{center}
\end{figure}

In order to derive the SZ flux of each cluster, we use an aperture photometry technique, consisting in integrating all pixels 
in a circle of radius $\theta_1$, and subtracting a background level calculated as the median of all pixels enclosed in an external 
ring between radii $\theta_2$ and $\theta_3$. 
For the aperture we use $\theta_1={\rm max}\left[\theta_{500},0.75\theta_{\rm FWHM}(\nu)\right]$, 
where $\theta_{500}$ is the angular radius where the mean density has dropped to 500 times the critical density at the cluster redshift, 
and $\theta_{\rm FWHM}(\nu)$ is the beam FWHM at a given frequency $\nu$. This ensures capturing most of the SZ flux for each cluster.
The bulk (typically $>60\%$, depending on the cluster profile) of the SZ flux is enclosed inside $\theta_{500}$. On the other hand, in order 
to avoid the possible loss of flux produced by the beam convolution, in those cases where $0.75\theta_{\rm FWHM}(\nu)>\theta_{500}$ 
we use $0.75\theta_{\rm FWHM}$ as the aperture radius. The external ring must not be too close to the cluster as this would result in an 
auto-subtraction of SZ flux, nor too far as it would not be representative of the true background around the cluster. We initially consider 
three cases: [$\theta_2$,$\theta_3$]=[$1.5$,$2.5$]$\theta_1$, [$2.0$,$2.24$]$\theta_1$ (in this case the aperture and the ring have equal 
areas) and [$3$,$4$]$\theta_1$. As shown in Figure~\ref{fig_3ringfluxes}, we get fluxes consistent at the $1\sigma$ level for these three cases.  
We finally decided to use the nearest ring, [$\theta_2$,$\theta_3$]=[$1.5$,$2.5$]$\theta_1$. In order to correct for the SZ flux that 
is lost when subtracting the background level, we calculate the SZ flux in the external ring using the universal profile from 
\cite{Arnaud2010} (we also tried the $\beta$-model profile and found no significant differences for the $T_{\rm CMB}(z)$ extraction), and add this value to our final flux 
estimate.  This correction factor is typically $\sim 5-20\%$.

The final flux is then given by 
\begin{equation}
S_{\rm SZ}(\nu) = \left[\frac{\sum_{i=1}^{n1} M(\nu)_{{\rm c},i}}{n_{\rm 1}} - {\tilde M(\nu)}_{{\rm c},j} \right] n_1 \Omega_{\rm pix} + S_{\rm SZ}^{\rm bg}~~,
\label{eq:ssz}
\end{equation}
where $n_1$ is the number of pixels in the aperture, and $M(\nu)_{{\rm c},i}$ and $M(\nu)_{{\rm c},j}$ represent respectively the pixel thermodynamic temperatures
in the aperture and in the background annulus, in the final cleaned maps. The median is calculated over the $n_2$ pixels in this annulus. 
$S_{\rm SZ}^{\rm bg}$ is the median SZ flux in the background estimated using a universal profile. $\Omega_{\rm pix}$ is the solid angle subtended 
by each pixel.

The calculation of the error associated to the flux derived through equation~\ref{eq:ssz} is crucial. It could be calculated analytically
as
\begin{equation}
\sigma(S_{\rm SZ}(\nu)) = \left[\frac{\sum_{i=1}^{n1}\sigma_i^2}{n_1^2}  + \frac{\pi}{2}\frac{\sum_{i=1}^{n2}\sigma_j^2}{n_2^2}\right]^{1/2} n_1 \Omega_{\rm pix}
= \sigma(M_{{\rm c},j})\left[\frac{1}{n_1}+\frac{\pi}{2}\frac{1}{n_2}\right]^{1/2} n_1 \Omega_{\rm pix}~~,
\label{eq:err_ssz}
\end{equation}
where $\sigma_{i,j}$ represents the error associated to each individual pixel, which is estimated through the pixel-to-pixel dispersion in the 
background annulus, $\sigma(M_{{\rm c},j})$. This equation however assumes perfect white noise (no pixel-to-pixel correlations). This would be 
a reliable assumption in the case of maps dominated by instrumental noise, but not in the case of non-negligible foreground residuals. 
We then chose a more conservative way of estimating the noise, consisting on calculating the flux via aperture photometry (with the same 
values of $\theta_1$, $\theta_2$ and $\theta_3$ as in the central position) on 10 evenly-separated positions located along a circle of radius $3\theta_3$. 
The final error is given by the standard deviation of the 10 calculated fluxes. We have checked that this estimate 
is consistent with equation~\ref{eq:err_ssz} for 70~GHz, where the foreground residuals are minimal. At higher frequencies the difference 
with respect to equation~\ref{eq:err_ssz} increases progressively, reaching a factor $\sim 3.5$ at 353~GHz. This indicates that the contribution 
of the foreground residuals to the noise might be $\sim 3$ times larger than the instrumental noise at this frequency.
The same 10 apertures around each cluster are used to determine the covariance matrix between different frequency bands. The cleaned maps at different frequencies may not be completely independent due to the possible presence of correlated foreground residuals and of common instrumental noise coming from the IRIS, 857 GHz, and LGMCA maps, that are used at all frequencies. We found that the off-diagonal terms of the normalized covariance matrix are typically below 0.5, significantly lower than what was obtained by \cite{Hurier2014}. The exact values depend strongly on the angular size of the cluster. 
We see higher correlation terms mainly for the two highest frequencies in our analysis (217 and 353 GHz) and only for clusters whose $R_{500}$ is greater than 8 arcmin. 

\section{\label{method_tvsz}Method for obtaining $T_{\rm CMB}$(z) measurements}

The possibility of determining $T_{\rm CMB}(z)$ from measurements of the
Sunyaev-Zel'dovich (SZ) effect was suggested long ago
\cite{fabbri1978, reph1980}. This effect -- Compton scattering of
the CMB by hot intracluster (IC) gas -- is a small change of the
CMB spectral intensity, $\Delta I_{\rm SZ}$, which depends on the
integrated IC gas pressure along the line of sight to the
cluster. The steep frequency dependence of the change in the CMB
spectral intensity, $\Delta I_{\rm SZ}$, due to the SZ effect allows the
CMB temperature to be estimated at the redshift of the cluster.
The spectral signature of the SZ, $\Delta I_{\rm SZ}$, depends on
frequency $\nu$ through the non-dimensional frequency x=$\frac{h\nu(z)}{kT(z)}=\frac{h\nu_0}{kT_0}$: it is redshift
invariant only for the standard scaling of $T(z)$. In all other non-standard scenarios,
the `almost' universal dependence of SZ on frequency becomes z-dependent,
resulting in a small dilation/contraction of the SZ spectrum on the frequency axis. 
The measured temperature change $\Delta T_{\rm SZ}$ of the CMB due to the SZ effect is given by  
\begin{equation}
	\Delta T_{\rm SZ}(x) = T_0 \tau [\theta f(x) -\beta   + R(x,\theta,\beta)]  
\end {equation}
where $\tau =\sigma_T \int n_e dl $ is the optical depth, $T_0$ is the CMB temperature at redshift $z=0$, 
$\theta=\frac{k_BT_e}{m_ec^2}$ with $T_e$ electron cluster temperature (we are assuming isothermality), $\beta= \frac{v_z}{c}$ with $v_z$ the radial component 
of the peculiar velocity of the cluster, and the $R(x,\theta,\frac{v_z}{c})$ function includes relativistic corrections \cite{yoel1995,itoh2002,Shimon2002}. 
If we assume that $T_{\rm CMB}$ scales with $z$ as $T_{\rm CMB}(z)=T_0(1+z)^{1-\beta}$, while the frequency scales as
$(1+z)$ as usual, then the non-dimensional frequency will be $x^{\prime}=\frac{h\nu_0}{k_B T^{\ast}_{\rm CMB}}$ and
$T^{\ast}_{\rm CMB}=T_{\rm CMB}(z)/(1+z)=T_0(1+z)^{-\beta}$ will be slightly
different from the local temperature $T_0$ as measured
by COBE-FIRAS. In this way it is possible to measure the temperature of
the CMB at the redshift of the cluster, thus directly 
constraining scenarios like those discussed in the introduction.  
Actually what we measure is the temperature change integrated over the solid angle occupied by the source in the sky, so we have:
\begin{equation}
	\Delta \mathfrak{T}_{\rm SZ} = \int \Delta T_{\rm SZ}(x) d\Omega = T_0 [\theta f(x) -\beta   + R(x,\theta,\beta)]\int\tau d\Omega~~. 
\end {equation}
In the following we will use $\mathcal{T} = \int\tau d\Omega $.
We use the SZ fluxes at Planck frequencies for each cluster in our sample.
In order to estimate the $\deltatomega(\nu_0)$ we need to evaluate the band integration of the SZ signal
\begin{equation}
	\deltatomega(\nu_0) = \int \deltatomega (\nu)t^{\prime}(\nu) d\nu
\end {equation}
where $t^{\prime}(\nu)$ is the spectral transmission.  
We use the band-average spectral transmission for each of the Planck frequency channels, as given in the RIMO (Reduced Instrument Model)\footnote{\url{http://wiki.cosmos.esa.int/planckpla/index.php/The_RIMO\#Effective_band_transmission_profiles}} and described in detail in \cite{cpp9}. 
Assuming that the SZ data are gaussianly distributed allows us to construct the single cluster likelihood function. The likelihood for the i-th cluster is:
\begin{equation}\begin{split}
	P(\Delta & \mathfrak{T}_{\rm obs}| \Theta_i)  \\
	  \propto &  \exp \left\{ -\sum_{jk} [\deltatomega(\nu_j,\Theta_i)-\Delta \mathfrak{T}_{\rm obs}(\nu_j)]Cov^{-1}(\nu_j,\nu_k)[\deltatomega(\nu_k,\Theta_i)-\Delta \mathfrak{T}_{\rm obs}(\nu_k)]\right \} 
\end{split}\end{equation}
where $\Theta_i =[\mathcal{T}_i,\theta_{i},\beta_i,T_{\rm CMB}(z_i),C]$, $Cov(\nu_j,\nu_k)$ is the covariance matrix between frequencies $\nu_j$ and $\nu_k$, and C is an uncertain scale factor, accounting for the calibration uncertainty.
The CMB temperature at the redshift of each cluster was extracted by performing statistical analysis on the observed thermodynamic temperature integrated over the solid angle occupied by the source in the j-th band. Here we apply one of the two direct methods developed in \cite{Luzzi2009}. 
	\subsection{\label{mcmc}MCMC analysis}
	
	The analysis has been performed through a Monte Carlo Markov Chain (MCMC) approach, which allows us to explore the full space of the cluster parameters (integrated optical depth $\mathcal{T}$, peculiar velocity $v_{pec}$, electron 	temperature $T_e$) and the CMB brightness temperature at the
	redshift of the cluster. In the analysis we allowed for calibration uncertainty, 
	an uncertain scale factor, C, modeled as a Gaussian with mean 1 and 1$\%$ standard
	deviation (calibration uncertainty is at a level of 0.5\% for the lower frequencies and 1\% at 353GHz \cite{cpp8}).
	The impact of absolute calibration accuracy will mainly influence the $\mathcal{T}$ parameter estimation, since $ \Delta \mathfrak{T} $ depends
linearly on $\mathcal{T}$ at the first order. 
	The MCMC generates random sequences of parameters, which simulate posterior distributions for all parameters 
	\cite{lewis2002}. The sampling approach we used is the one proposed by Metropolis and Hastings
	\cite{metropolis1953, hastings1970}. We run three chains for each cluster; convergence and mixing of the MCMC runs was 
	tested through the Gelman-Rubin test \cite{gelman1992}. 
	We included gaussian prior over the cluster gas temperature, as provided by X-ray data when available, or as specified in ~\autoref{subPSZ}.
	
	As noted in ~\autoref{cleaning} we clean our frequency maps from CMB. Kinematic SZ (KSZ) and the
	CMB primary anisotropy have the same spectral shape in the non relativistic limit, thus in cleaning from CMB we have also removed the kinematic component of the SZ. 
	In order to take into account possible CMB and KSZ residuals after the removal we model them as a kinematic SZ component and we adopt as a prior a Gaussian with a universal vanishing mean and with a $500$ km/s standard 		deviation. This corresponds to a residual CMB signal of $\sim 50 \mu \rm K$, which is compatible with the expected residual level of the LGMCA map \cite{bobin14}. 
	As was already noted in \cite{Luzzi2009} there is a degeneracy between $T_{\rm CMB}(z)$ and $v_{pec}$. In order to reduce the impact of this degeneracy and then to reduce the uncertainty
in the determination of $T_{\rm CMB}$(z), a better knowledge of the peculiar velocity is required or
it is necessary to remove the kinematic component from the thermal component, together
with the CMB primary anisotropy. Even if after the cleaning we are left with only a residual kinematic component this prevented us from obtaining ~$1\%$ sensitivity on $T_{\rm CMB}$(z) measurements, as forecasted in \cite{Avgoustidis2012}. 
	For the integrated optical depth $\mathcal{T}$ we use a very broad 	gaussian prior, which is centered on the Planck value of the Compton parameter at $R_{500}$ from SZ-proxy $\rm Y_{z}$ (after converting it in the cylindrical quantity),  divided by the $T_e$ value from X-rays, and with standard deviation 			equal to its central value. 
	
	As already specified in \autoref{aph}, we have adopted the universal pressure profile of \cite{Arnaud2010} in our analysis, nevertheless we have also verified that assuming the isothermal $\beta$ model \cite{Cavaliere1978} produces almost no 	difference with respect to the final $T_{\rm CMB}(z)$ extraction, the only difference being in the $\mathcal{T}$ determination.

	For the $T_{\rm CMB}(z)$  we allow for a broad gaussian prior centered on the standard value $T_0(1+z)$ and with standard deviation $0.5\rm K(1+z)$. We will show that this has no impact on the final precision with which we extract 		$T_{\rm CMB}(z)$ but prevents the chains from spending much time exploring unphysical values for this parameter. 
	The need for a good knowledge of $T_e$ is mainly dictated by the fact that we want to use the same cluster subsample (i.e. use the extracted $\mathcal{T}$) to build the Hubble diagram and do distance duality tests. Having a narrow 		prior on $T_e$ has as first outcome that $\mathcal{T}$ is better constrained, moreover the MCMC converge more quickly with respect to the case in which the prior on $T_e$ is broad.
	
	In our analysis we have included the relativistic expression $R(x,\theta,\beta)$, we used the formulation given in \cite{itoh2002}. To give a hint of the effect of relativistic corrections we report results for $T_{\rm CMB}(z)$ for the two 		highest z clusters in our sample. 	
	ACT-CLJ0102-4915 (El Gordo) is at z=0.87, $T_X = 14.5 \pm 0.1 \rm keV$ \cite{Menanteau2012}, with relativistic corrections we have $T_{\rm CMB}(z) = (4.85 \pm 0.25)\rm K$ (5.096 K standard value),  without 	relativistic corrections $T_{\rm CMB}(z) = (4.99 \pm 0.27)\rm K$. For PSZ1 G266.56-27.31, with z=0.97, $T_X\sim 11 \rm keV$ \cite{pip26}, with relativistic corrections we have $T_{\rm CMB}(z) = (5.32 \pm 0.27)\rm K$ (5.374 K standard value), without relativistic corrections $T_{\rm CMB}(z) = (5.44 \pm 0.27)\rm K$. Both values are consistent within the error, still using relativistic corrections (when $T_e$ is relatively high) changes the $T_{\rm CMB}(z)$ determination by $\sim 2 \%$.
	
\begin{figure}
	 \begin{center}
    		\includegraphics[width=16cm,keepaspectratio]{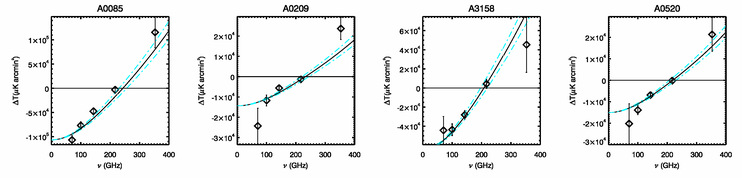}\\
		\includegraphics[width=16cm,keepaspectratio]{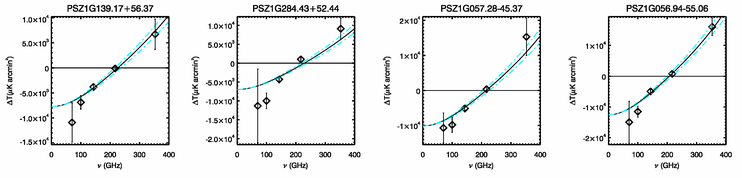}\\
		\includegraphics[width=16cm,keepaspectratio]{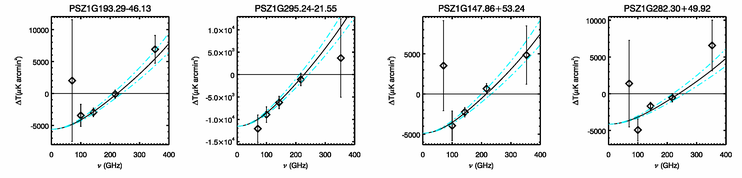}\\
		\hspace*{3.9cm}
		\includegraphics[width=16cm,keepaspectratio]{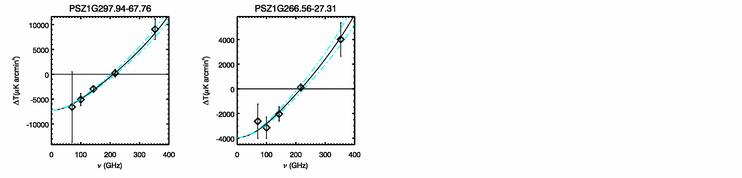}\\
    		\caption{\label{fig_spectra} 
                 Measured SZ fluxes.  Solid lines are the best fit spectra, cyan dash dotted lines are 1$\sigma$ errors on the $T_{\rm CMB}(z)$ parameter;  these plots show the importance of including high frequency measurements for $T_{\rm CMB}(z)$ extraction. From top to bottom: each row corresponds to the various cluster subsamples, selected with the criteria explained in~\ref{subPSZ}. Clusters in each row have SNR on $T_{\rm CMB}(z)$ corresponding to the average SNR in the specific subsample.}
  	 \end{center}
	\end{figure}

In Fig. \ref{fig_spectra} we show the SZ fluxes and best fit spectra for a selection of our cluster sample. 

\section{\label{results}Results}

	\subsection{\label{beta} Cosmological constraints}	
\begin{figure}
	 \begin{center}
    		\includegraphics[width=16cm,keepaspectratio]{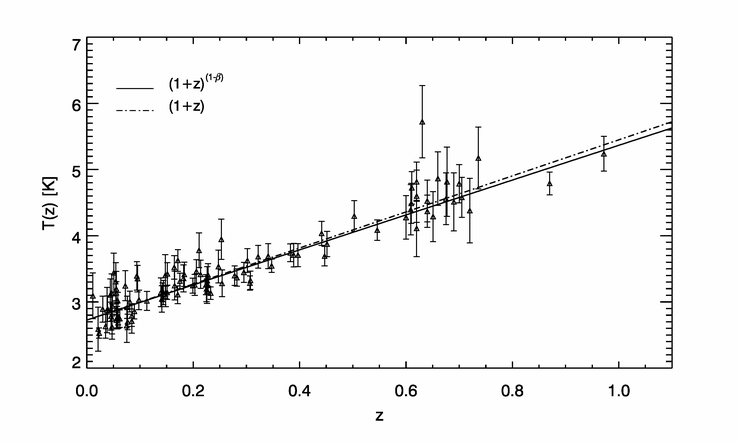}
    		\caption{\label{fig_tvsz} 
                 Measurements of the CMB temperature as a function of redshift for 104 Planck selected clusters. The solid
line is the best fit to the scaling $T_{\rm CMB}$(z)=$T_0$(1+z)$^{1-\beta}$. The dot-dashed line is the standard scaling, with $\beta=0$.}
  	 \end{center}
	\end{figure}

\begin{figure}
	 \begin{center}
    		\includegraphics[width=12cm,keepaspectratio]{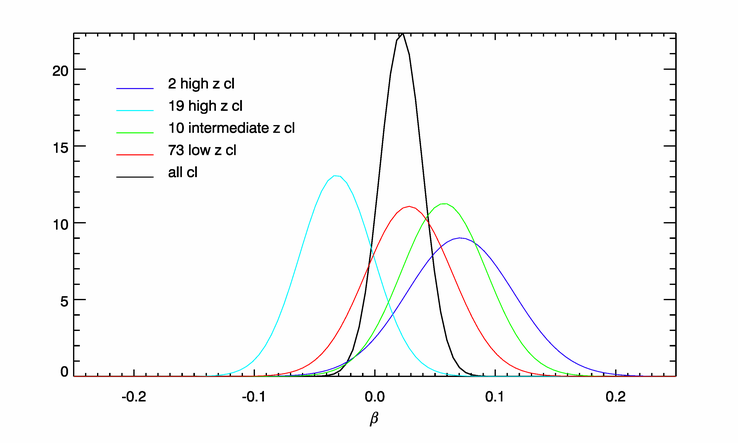}
    		\caption{\label{fig_betaposterior} 
                 Posterior of the $\beta$ parameter as derived from the joint pdf on our cluster sample. The black-line is the posterior on the full cluster sample, colored lines show the joint pdf for each cluster subsample selected as in~\ref{subPSZ}. }
  	 \end{center}
	\end{figure}

	In fig. \ref{fig_tvsz} we plot the measurements of the CMB temperature as a function of redshift. We report expected values and standard deviation of each single $T_{\rm CMB}$(z) distribution.  
	Our measurements reach a precision of up to $3\%$ and of $7\%$ on average on 104 clusters. Of our initial sample of 107 clusters, Abell0773 and Abell1767 didn't converge, PSZ1G205.94-39.46 is an outlier; a more extended analysis on this cluster will be discussed in a follow-up publication.  												
	Since the distributions of $T_{\rm CMB}(z)$ for individual clusters are in general slightly
	non-Gaussian and in addition they are frequently skewed, performing a best fit
	as if they were Gaussian can introduce a bias in the result. 
	In order to limit possible deviations of the form $T_{\rm CMB}$(z)=$T_0$(1+z)$^{1-\beta}$ we have taken the full posteriors of $T_{\rm CMB}(z)$ at each redshift and we derive the joint pdf for the parameter $\beta$.
	The posterior for $\beta$ is shown in fig. \ref{fig_betaposterior}.  The bestfit result we get is: $\beta$ = $0.022 \pm 0.018$, at 1$\sigma$ uncertainty, in agreement with the predictions of the standard model. Our result, obtained with individual determinations of $T_{\rm CMB}$(z) at each cluster's redshift, on 104 Planck selected clusters,  is compatible with and at the same level of precision of the result from \cite{Hurier2014}, which found on 813 Planck stacked clusters $\beta= 0.009 \pm 0.017$. 
	Also, our result is compatible  with and at a higher level of precision of the result from \cite{Saro2014}, which found on 158 SPT-selected clusters $\beta=0.017 ^{+0.030}_{-0.028}$. 
	Assuming the standard evolution of the CMB temperature with redshift we can use our measurements of $T_{\rm CMB}$(z) to put constraints on the value of $T_{\rm CMB}$ at redshift zero. We get: $T_{\rm CMB}(0)= 2.702 \pm 0.013~ \rm K$, at 1$\sigma$ uncertainty, value consistent with the COBE-FIRAS measurement, $T_0=2.7260 \pm 0.0013 ~\rm K$ \cite{Fixsen2009}.

\subsection{Combined constraints with previous measurements}
By combining our results with previous result from SZ measurements \cite{Saro2014} and from spectroscopic studies
of lines in absorption against quasars \cite{Ge1997,Srianand2000,Molaro2002,Cui2005,Noterdaeme2011,Muller2013}, we obtain a tighter constraint
on $\beta$, finding $\beta = 0.016 \pm 0.012$. 
Our combined results are slightly less compatible with the standard model, still they are compatible with the constraints found by \cite{Saro2014}, $\beta=0.005 \pm 0.012$, combining their SPT results with those from spectroscopic measurements and from 813 Planck stacked clusters from \cite{Hurier2014}.  
We could not include the result from \cite{Hurier2014}, $\beta= 0.009 \pm 0.017$,  since it was obtained from 813 stacked Planck clusters, and this sample includes also the 104 clusters we have used for our analysis. Although the analyses are completely different there is a redundancy of measurements on some cluster: it is not possible for us to isolate from the constraints of \cite{Hurier2014} the contribution from our sample. 

We also note that, most probably because of the larger spectral coverage of Planck data with respect to SPT data, with the method we adopted to provide individual determinations of $T_{\rm CMB}$(z) at each cluster's redshift we can provide stronger constraints on a per cluster basis and on the overall $\beta$, this even though the 158 SPT-selected clusters in \cite{Saro2014} span a larger redshift range.

\begin{figure}
	 \begin{center}
    		\includegraphics[width=16cm,keepaspectratio]{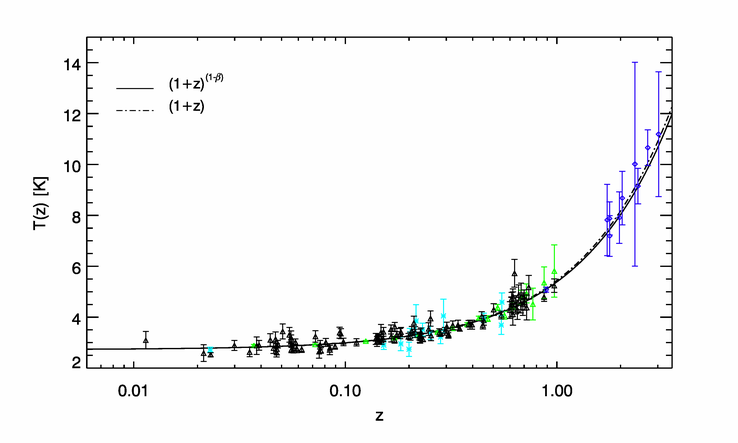}
    		\caption{\label{fig_tvsz_all} 
                 Measurements of the CMB temperature as a function of redshift. Black points are measurements  from SZ spectra of individual clusters (104 Planck-selected clusters); blue points are absorption lines measurements (Muller et al 2013 and references therein); T(z) data points from Saro et al. (2014) are not available.
		The solid line is the best fit to the scaling $T_{\rm CMB}$(z)=$T_0$(1+z)$^{1-\beta}$ obtained by combining our result with Muller et al. (2013) and with Saro et al. (2014). The dot-dashed line is the standard scaling. For visualization purpose we also plot in cyan previous SZ measurements toward galaxy clusters (Luzzi et al. 2009) and in green stacked Planck SZ selected clusters (Hurier et al. 2014). We haven't used these two last datasets for our combined constraints on $\beta$, due to common clusters in our sample.}
  	 \end{center}
	\end{figure}
	
\section{\label{conclusions}Conclusions}
We have obtained reliable SZ spectra in the frequency range 70-- 353 GHz for a subsample of the Planck SZ cluster catalog with known X-ray properties. Taking advantage of the multifrequency measurements of the SZ effect and of the high quality of data, we have obtained individual determinations of $T_{\rm CMB}$(z) at cluster redshift for 104 clusters, with a precision of up to $3\%$ and on average of $7\%$ on the full sample. 
We have shown that relativistic corrections have an impact on the determination of $T_{\rm CMB}(z)$ for clusters with high electronic temperature and when including high frequencies, the latest much needed for this study.  
We have studied possible deviations of the form $T_{\rm CMB}$(z)=$T_0$(1+z)$^{1-\beta}$, getting the constraint $\beta$ = $0.022 \pm 0.018$. Our results are in agreement with the predictions of the standard model. They are compatible with and at the same level of precision of previously
published results based on galaxy clusters \cite{Hurier2014,Saro2014}  and on spectroscopic studies
of lines in absorption against quasars \cite{Muller2013}. Our results are compatible with those obtained by \cite{deMartino2015} using a similar foreground-cleaning methodology, and a larger sample of 481 X-ray selected clusters out to $z=0.3$. Given the strong connection between cosmological distance duality relation and the CMB temperature scaling law \cite{Avgoustidis2012} we plan to use 86 of the SZ spectra obtained in this work (clusters with known R500) to investigate the distance duality relation applying the methods proposed by \cite{Uzan2004,Holanda2012}.
We work to extend our analysis to a larger sample, namely the 813 Planck confirmed clusters with known redshift, improving the level of homogeneity of the available X-ray informations.
In the longer term, a COrE-like experiment \cite{PRISM2014}, with extended  frequency coverage with respect to Planck, will bring significant further improvements.

\section*{Acknowledgements}

We acknowledge the use of the HEALPix package. We acknowledge the use of data from the Planck/ESA mission, downloaded from the Planck Legacy Archive.\\ 
This research has made use of the X-Rays Clusters Database (BAX) which is operated by the Laboratoire d'Astrophysique de Tarbes-Toulouse (LATT), under contract with the Centre National d'Etudes Spatiales (CNES). This research has made use of the SZ-Cluster Database operated by the Integrated Data and Operation Center (IDOC) at the Institut d'Astrophysique Spatiale (IAS) under contract with CNES and CNRS.
This work was done in the context of project PTDC/FIS/111725/2009 (FCT, Portugal). CJM is also supported by an FCT Research Professorship, contract reference IF/00064/2012, funded by FCT/MCTES (Portugal) and POPH/FSE (EC).
Our analysis extends work developed in the cooperation grant `Probing
Fundamental Physics with Planck' (PHC-EGIDE/PESSOA, FCT/1562/25/1/2012/S). This work has also been partially 
supported by funding from the University of Rome Sapienza Anno 2013 - prot. C26A135JC3.

\end{document}